\documentclass[a4paper]{PoS}

\usepackage{lineno}
\usepackage{float}
\usepackage{subcaption}

\vspace{-10cm}

\title{Search for electromagnetic super-preshowers using gamma-ray telescopes}

\ShortTitle{Search for electromagnetic super-preshowers using gamma-ray 
telescopes}

\abstract{ \footnotesize Any considerations on propagation of particles through the Universe 
must involve particle interactions: processes leading to production of 
particle cascades. While one expects existence of such cascades, the state of 
the art cosmic-ray research is oriented purely on a detection of single 
particles, gamma rays or associated extensive air showers. The natural 
extension of the cosmic-ray research with the studies on ensembles of particles 
and air showers is being proposed by the Cosmic-Ray Extremely 
Distributed Observatory (CREDO) Collaboration. Within the CREDO strategy the 
focus is put on generalized super-preshowers (SPS): spatially and/or temporally 
extended cascades of particles originated above the Earth atmosphere, 
possibly even at astrophysical distances. With CREDO we want to find out 
whether SPS can be at least partially observed by a network of terrestrial 
and/or satellite detectors receiving primary or secondary cosmic-ray signal. 
This paper addresses electromagnetic SPS, e.g. initiated by very high 
energy photons interacting with the cosmic microwave background, and the SPS 
signatures that can be seen by gamma-ray telescopes, exploring the example
of Cherenkov Telescope Array. The energy spectrum of secondary electrons and 
photons in an electromagnetic super-preshower might be extended over a
wide range of energy, down to TeV or even lower, as it is evident from the 
simulation results. This means that electromagnetic showers induced by such
particles in the Earth atmosphere could be observed by imaging atmospheric 
Cherenkov telescopes. We present preliminary results from the study of 
response of the Cherenkov Telescope Array to SPS events, including the analysis 
of the simulated shower images on the camera focal plane and implemented
generic reconstruction chains based on the Hillas parameters. The eventual 
detection of super-preshowers will mean opening a new window to the Universe
- a new chance for solving the most pending issues of contemporary 
astrophysics.}

\FullConference{35th International Cosmic Ray Conference -- ICRC2017-\\
		10-20 July, 2017\\
		Bexco, Busan, Korea
		}

\author{K. Almeida Cheminant$^{*1}$, D. G\'{o}ra$^1$, N. Dhital$^1$, P. Homola$^1$, P. Pozna\'{n}ski$^2$, \L{}. Bratek$^1$, T. Bretz$^3$, P. Jagoda$^{1,\ 4}$, J. Ja\l{}ocha$^2$, J. Jarvis$^{1,\ 5}$, K. Kopa\'{n}ski$^1$, M. Krupinski$^1$, D. Lema\'{n}ski$^{1,\ 2}$, V. Nazari$^{1,\ 6}$, J. Niedzwiedzki$^4$, M. Nocu\'{n}$^4$, W. Noga$^1$, A. Ozieblo$^7$, K. Smelcerz$^{1,\ 2}$, K. Smolek$^8$, J. Stasielak$^1$, S. Stuglik$^{1,\ 2}$, M. Su\l{}ek$^{1,\ 2}$, O. Sushchov$^1$ and J. Zamora-Saa$^6$
\\* speaker: \textit{E-mail}: \email{kevin.almeida-cheminant@ifj.edu.pl}\\
\footnotesize
\llap{$^1$}Institute of Nuclear Physics Polish Academy of Sciences, Radzikowskiego 152, Cracow, Poland\\
\llap{$^2$}Cracow University of Technology, Warszawska 24, Cracow, Poland\\
\llap{$^3$}RWTH Aachen University, Physics Institute III A, Aachen, Germany\\
\llap{$^4$}AGH University of Science and Technology, al. Mickiewicza 30, 30-059 Cracow, Poland\\
\llap{$^5$}School of Physical Sciences, Open University, Buckinghamshire, MK7 6AA, United Kingdom\\
\llap{$^6$}Joint Institute for Nuclear Research, Dubna 141980, Russia\\
\llap{$^7$}AGH University of Science and Technology, ACC Cyfronet AGH, ul. Nawojki 11, 30-950 Cracow, Poland\\
\llap{$^8$}Institute of Experimental and Applied Physics, Czech Technical University in Prague, Horsk\'{a} 3a/22, 128 00 Praha 2, Czech Republic}

\begin{document}

\section{Introduction}
\vspace{-0.4cm}
Two mysteries of contemporary astroparticle physics revolve around the nature of dark matter and the existence of cosmic rays (CRs) with energies greater than 
$10^{20}$eV. A possible answer to these questions can be found in the Super Heavy Dark Matter (SHDM) scenario \cite{chung99}. It assumes a production of supermassive (i.e. $M_{X} > 10^{23}$eV) particles in the early Universe, during the inflation phase. Such particles could presently annihilate or decay leading to the production of jets containing mainly photons \cite{berezinsky97}. The energy of these photons could easily be of the order of $10^{20}$eV, the value that seems to be out of reach with standard astrophysical acceleration mechanisms. The key prediction of this scenario in the SHDM group is that the ultra-high energy cosmic-ray (UHECR) flux observed at the Earth should be dominated by photons \cite{rubtsov06}. On the other hand, the highest energy events observed by the leading collaborations: Pierre Auger\footnote{www.auger.org} and Telescope Array\footnote{www.telescopearray.org}, are not considered photon candidates if the present state-of-art air shower reconstruction procedures are applied. In fact there are no photon candidates also in the data below $10^{20}$eV collected by these and other observatories, leading to very stringent upper limits \cite{niechciol17} (see Figure 1 - left panel). However, two main doubts about such conclusions can be addressed: a) the present state of the art analysis does not take into account mechanisms that could lead to a good mimicking of hadronic air showers with the showers induced by photons, and b) the present state of the art analysis does not take into account mechanisms that could lead to the efficient screening/cascading of the ultra-high energy (UHE) photons on their way to the Earth, so that the products of such screening/cascading are out of reach of the presently operating observatories, which is interpreted as non-observation of UHE photons. A deeper understanding of the upper limits to UHE photons require an awareness of the assumptions underlying this result. It is also important to understand that constraints on dark matter models or fundamental physics laws can be inferred from the upper limits to UHE photons only assuming no unexpected propagation effects occurring to the UHE photons on the way from the production site to the Earth. Let's assume a hypothetical mechanism leading to a cascading of most of the UHE photons before they reach Earth, leading to the efficient shrinking of their astrophysical horizon. If such a mechanism or process occurs in reality, UHE photons have little chance to reach the Earth, and what can be observed on Earth is the result of the mentioned underlying hypothetical mechanism, most likely large electromagnetic cascade formation. The Cosmic-Ray Extremely Distributed Observatory (CREDO) \cite{homola17} focuses on the observation of these cosmic cascades by a global network of CR detectors. One example of such a cascading process is the preshower effect \cite{mcbreen81} describing an interaction of a UHE photon and secondary electrons with the geomagnetic field, above the atmosphere and before the extensive air showers (EAS) are initiated by the preshower particles. A super-preshower (SPS) is a cascade of electromagnetic particles originated above the Earth atmosphere, no matter the initiating process. SPSs can be classified with respect to their principal observable properties: spread in space $\Delta$x and time $\Delta$t, as shown in Figure 1 - right panel. A cascade initiated by the interaction of a UHE photon with the geomagnetic field is contained within a few square centimeters (cases A and B). Simulations of such cascades are performed by the PRESHOWER program \cite{homola05} and the following work is constrained to these two cases. If the preshower effect would occur in the vicinity of the Sun, one would expect similarly negligible $\Delta$t, but $\Delta$x should be much larger that in case of SPS type A, maybe even close to the size of the Earth (classes C and D).

\begin{figure}[t]
\vspace{-0.7cm}
\begin{subfigure}{.49\textwidth}
  \centering
  \includegraphics[height=5cm]{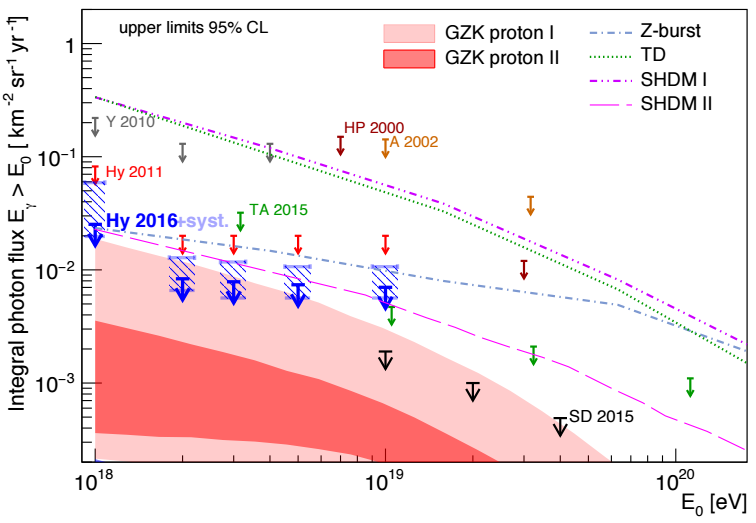}
  \label{fig:sfig1}
\end{subfigure}
\begin{subfigure}{.49\textwidth}
  \centering
  \includegraphics[width=7.4cm]{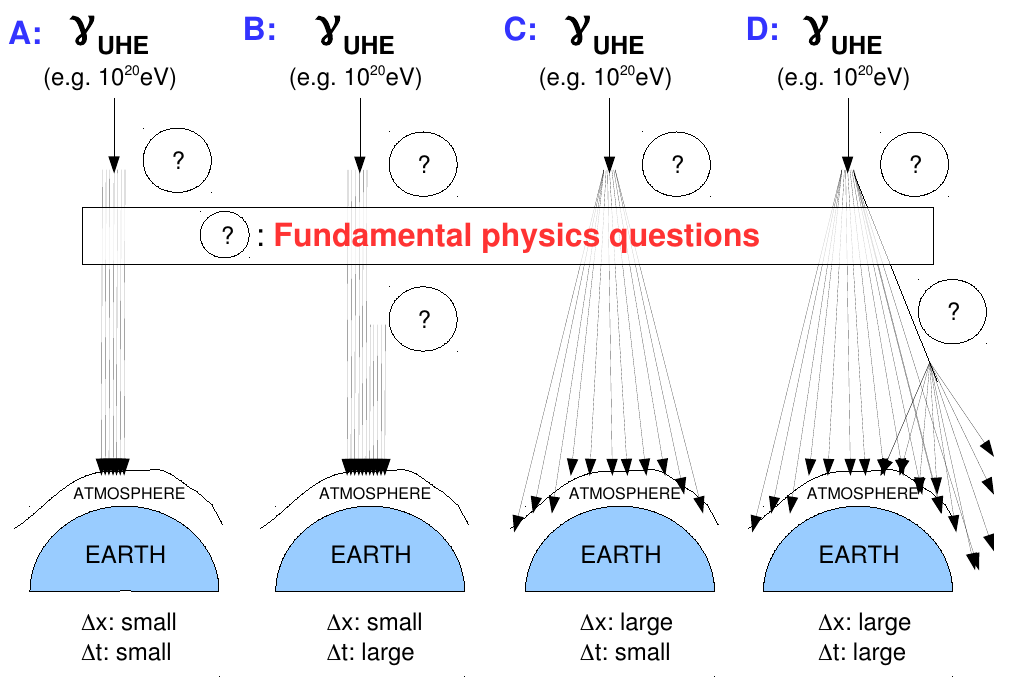}
  \label{fig:sfig2}
\end{subfigure}
\vspace{-0.3cm}
\caption{\textit{Left panel:} Upper limits at 95\% C.L. to the diffuse flux of UHE photons from different observatories and predictions from several top-down, and cosmogenic photon models. Taken from \cite{niechciol17}. \textit{Right panel:} A basic observational classification of SPSs. Different classes refer to different widths of spatial and temporal distributions of SPS particles. The question marks represent the uncertainties about the fundamental physics processes at $E = 10^{20}$eV or larger.}
\label{Figure 1}
\vspace{-0.4cm}
\end{figure}

In the present study we focus on the potential observation of SPSs by Imaging Atmospheric Cherenkov Telescopes (IACTs), which can detect the Cherenkov emission of charged particles composing the air showers generated by the preshower cascade. More specifically, the simulations were performed for the northern hemisphere site (La Palma) of the next-generation ground-based gamma-ray telescope, Cherenkov Telescope Array (CTA) \cite{cta11}. The simulated CTA array consists of 19 telescopes (4 Large-Sized Telescopes and 15 Medium-Sized Telescopes) as shown on the left panel of Figure 2. In this work, a special attention is given to nearly horizontal showers: although muons are less abundant near the shower maximum compared to other charged particles, they propagate over longer distances, leading to a predominance of the muonic component of the air shower over the electromagnetic one and to Cherenkov emission build-up that can be detected by IACTs at inclined directions i.e. above 70 degrees zenith angle. The abundance of these muons being directly dependent on the nature of the primary particle, the amount of detected Cherenkov light can help us discriminate between different primaries as shown in \cite{neronov16}. Using a similar method, we aim at discriminating cosmic-ray background and SPSs. As a direct consequence, detecting the muon signal of such inclined showers could lead to the improvement of the rejection of CR background in gamma-ray observations. The IACTs technique applied to the observation of high-zenith angle showers could therefore lead to a better sensitivity (two orders of magnitude) in the energy band above the knee compared to the KASKADE-Grande experiment \footnote{web.ikp.kit.edu/KASCADE} and about one order of magnitude in the energy range above the ankle compared to the Pierre Auger experiment \cite{neronov16}. 
\vspace{-0.4cm}
\section{Method}
\vspace{-0.3cm}
Simulating SPSs and studying their expected signature is a three-step process. First, the propagation of a photon several thousands of kilometers above sea level is simulated with the PRESHOWER algorithm \cite{homola05}. Because high-energy photons can produce an electron/positron pair when interacting with local magnetic field, conversion probability is checked along the  propagation of the primary photon in steps of 10 kilometers. Figure 2 (right) shows the conversion probability for different zenith angles in the case of 40 EeV primary photon coming at the La Palma site. If no conversion occurs, the simulation is terminated. Otherwise, the resulting electron/positron pair can interact with the Earth's magnetic field and generate bremsstrahlung radiation. The geomagnetic field value is taken for the location of the studied CTA array, i.e. the La Palma site. The resulting electromagnetic particles are treated as one shower by the air-shower simulation software CORSIKA \cite{heck98}. To maximize the chance of pair production, simulations were performed for zenith angle $\theta=80^{\circ}$ and azimuth angle $\phi=180^{\circ}$, i.e. when the maximum conversion probability is expected for fixed energy of the primary particle. In addition to PRESHOWER, CORSIKA was compiled with several other options. QGSJETII \cite{ostapchenko06} and URQMD \cite{bass98} were selected to model high and low energy hadronic interactions, respectively. To take into account Cherenkov emission from ultra-relativistic charged particles, both CERENKOV and IACT options were chosen \cite{bernlohr08}. Because simulations were performed for very inclined showers, Earth's atmosphere curvature had to be taken account and the CURVED option was activated. Finally, in order to investigate the detection sensitivity at different impact distances, the CSCAT option was selected, allowing the randomization of the impact point of the simulated shower in a circle of radius $R_{imp}$ defined on a plane perpendicular to the shower axis.

\begin{figure}[t]
\vspace{-0.7cm}
\begin{subfigure}{.49\textwidth}
  \centering
  \includegraphics[width=7cm]{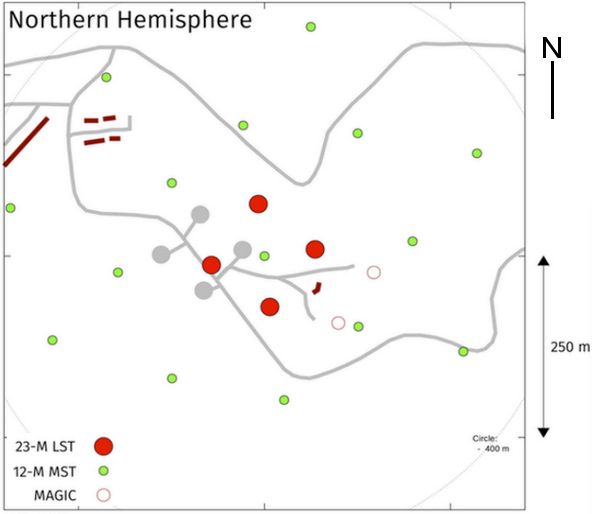}
  \label{fig:sfig20}
\end{subfigure}
\begin{subfigure}{.5\textwidth}
  \centering
  \includegraphics[height=5cm]{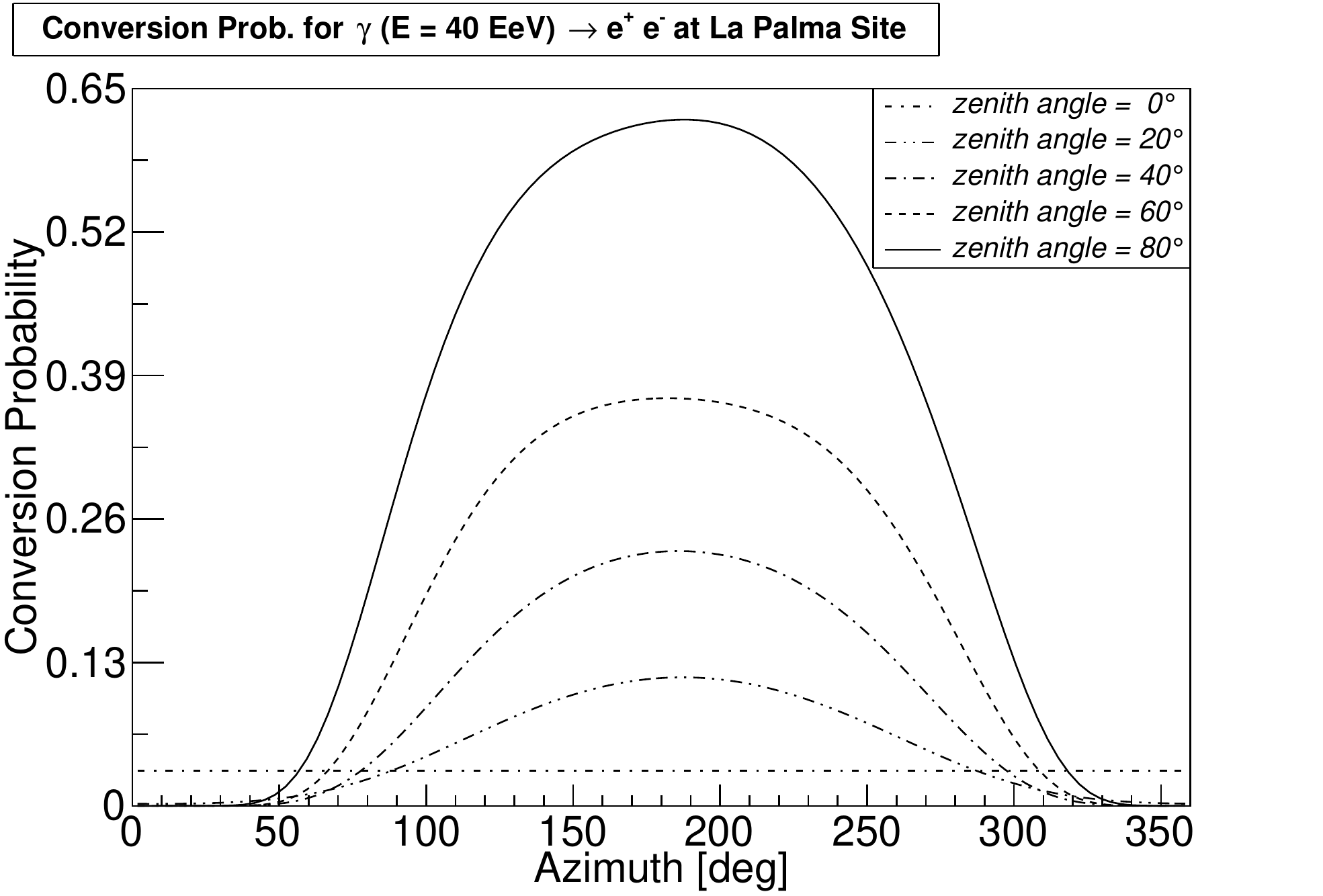}
  \label{fig:sfig21}
\end{subfigure}
\caption{\textit{Left panel}: Geometry of the simulated CTA array considered for the La Palma site taken from \cite{ctasite}. \textit{Right panel}: Conversion probability of 40 EeV photon into electron/positron pair for different zenith angles at La Palma site in CORSIKA frame of reference.}
\label{Figure 2}
\vspace{-0.4cm}
\end{figure}

The CORSIKA output was then piped into \textit{sim\_telarray} \cite{bernlohr08}, which is a software package for simulation of IACT detector response to Cherenkov photons. The package implements the detailed properties of the  detectors including the camera electronics response and the optical ray-tracing. The detector's response simulations were performed with the \textit{production-I} settings which takes into account the properties of the different types of CTA telescopes \cite{cta13}.A similar approach was taken to simulate showers generated by protons and photons without preshower effect. The VIEWCONE option was selected to vary the primary's arrival direction on a circle of aperture $\alpha=5^{\circ}$ perpendicular to the primary fixed direction. In this way we mimic the diffuse signal from photons and CRs.
\vspace{-0.4cm}
\section{Results}
\vspace{-0.3cm}
\begin{figure}[t]
\vspace{-0.7cm}
\begin{subfigure}{.49\textwidth}
  \centering
  \includegraphics[width=7.4cm]{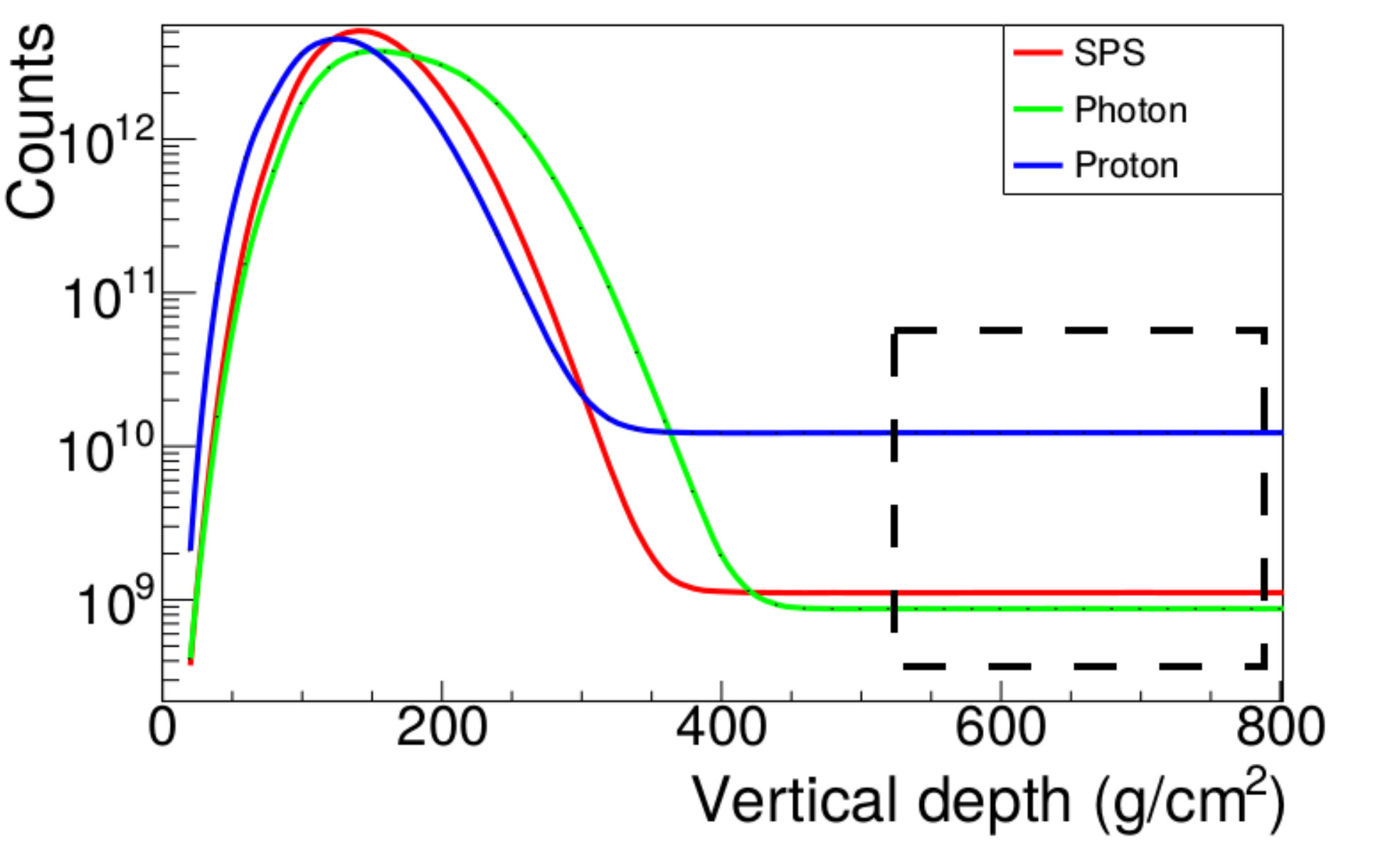}
  \label{fig:sfig11}
\end{subfigure}
\begin{subfigure}{.5\textwidth}
  \centering
  \includegraphics[width=7.4cm]{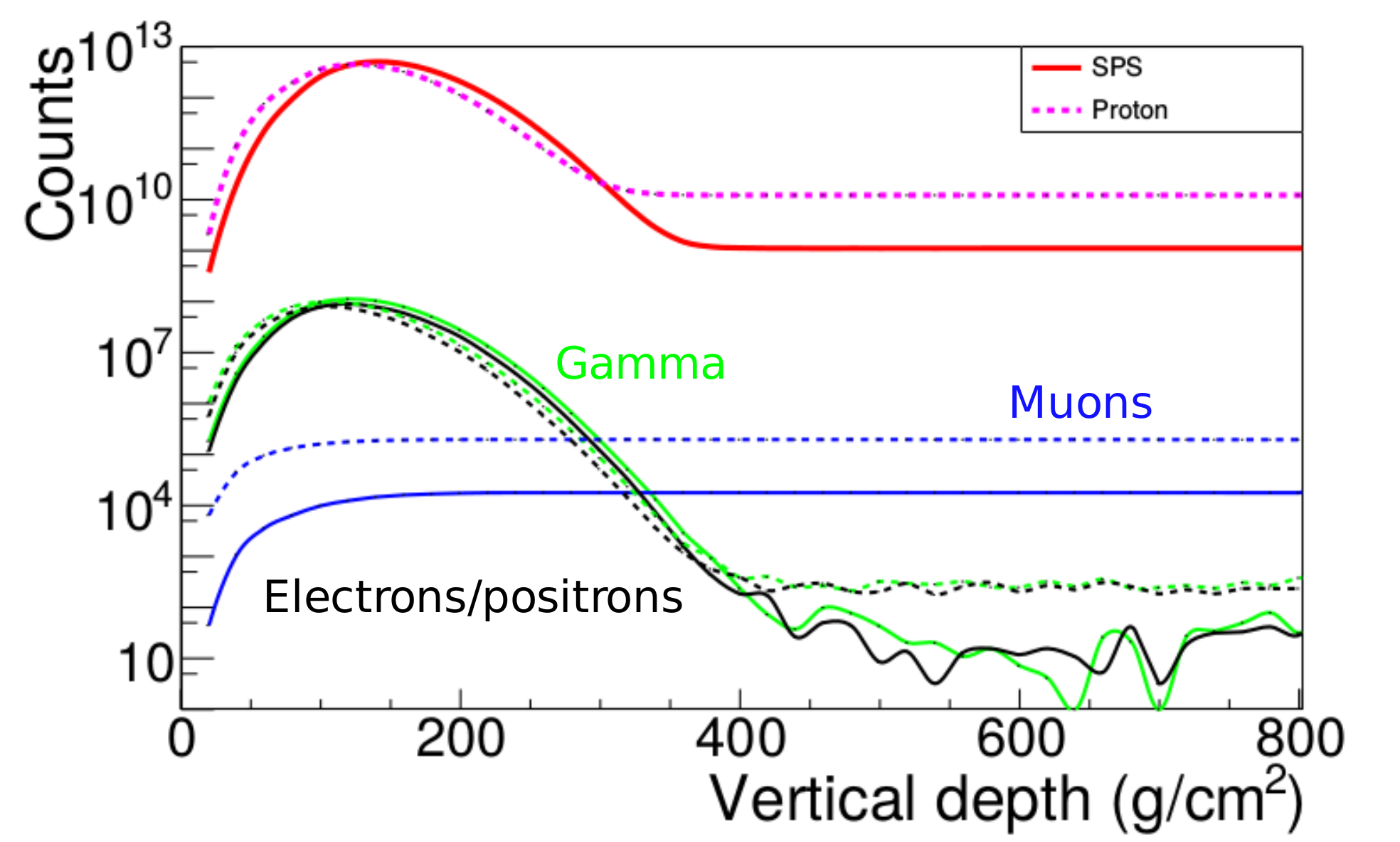}
  \label{fig:sfig12}
\end{subfigure}
\vspace{-0.3cm}
\caption{\textit{Left panel}: Cherenkov profile for different primaries of energy $E = 40$ EeV with direction $\theta=80^{\circ}$ and $\phi=180^{\circ}$ showing the different regimes of the shower development. Dashed black rectangle shows the part of the shower (dominated by muons) seen by IACTs when observing at large zenith angle. \textit{Right panel}: Cherenkov profiles of proton (pink) and SPS-induced (red) air-showers. Blue, green and black lines show the different components of the air-shower for the proton (dashed lines) and the SPS (solid lines) primaries.}
\label{Figure 3}
\vspace{-0.5cm}
\end{figure}

The longitudinal profiles plotted on the left panel of Figure 3 show the amount of Cherenkov light emitted by different primaries, of same injection parameters, as a function of the vertical atmospheric depth. For similar first interaction points, photon-induced showers reach their maxima deeper in the atmosphere than proton induced showers due to LPM effect \cite{lpm53}, which explains the shift of Cherenkov profiles of the former towards larger atmospheric depths. However, in the case of SPS (which can be treated as showers generated by multiple photons), the interaction point is much higher in the atmosphere and its maximum becomes slightly shifted to lower depths, towards proton-induced shower maximum. Consequently, observatories such as Pierre Auger for example could hardly discriminate between proton and SPS-induced air-showers with its current observation mode. In fact, SPS might have already been detected by different experiments but identified as protons. The main difference between these two types of shower lies in the Cherenkov emission after the shower maximum is reached, which is about one order of magnitude more important in the case of the proton primary. This part of the air-shower is dominated by the muon component, as shown in the right panel of Figure 3 and can be explained by the fact that the main channel of muon production is the decay of charged pions produced by hadronic interactions, absent in the case of photon-induced showers. Observing at large zenith angles would make detecting the muon plateau (Figure 3 - left panel, black rectangle) possible and therefore allow the potential identification of SPS. However, it is important to mention that above $10^{7}$ GeV, the uncertainties in the hadronic interaction increase with the energy \cite{risse06} and can lead to the different muon content. Therefore, for some models, the muon content of SPS-induced air showers could mimic the muon content and the longitudinal profile of extensive air showers (EAS) induced by nuclei.

Electrons suffer from high energy losses mostly due to ionization and bremsstrahlung radiation. However, they largely dominate the early development of the air-shower for proton, photon and SPS primaries, as shown on the right panel of Figure 3. Therefore, the Cherenkov emission peak is mostly caused by the electromagnetic component reaching its maximum. Muons, on the other hand, suffer energy losses to a lesser extent and can propagate over longer distances. As the shower develops, the muon content increases until it exceeds the amount of electron/positron. As the number of muons remains constant throughout the post-maximum shower development, the Cherenkov emission follows the same trend. 

\begin{figure}[t]
\vspace{-0.6cm}
\begin{subfigure}{.49\textwidth}
  \centering
  \includegraphics[width=.5\textwidth]{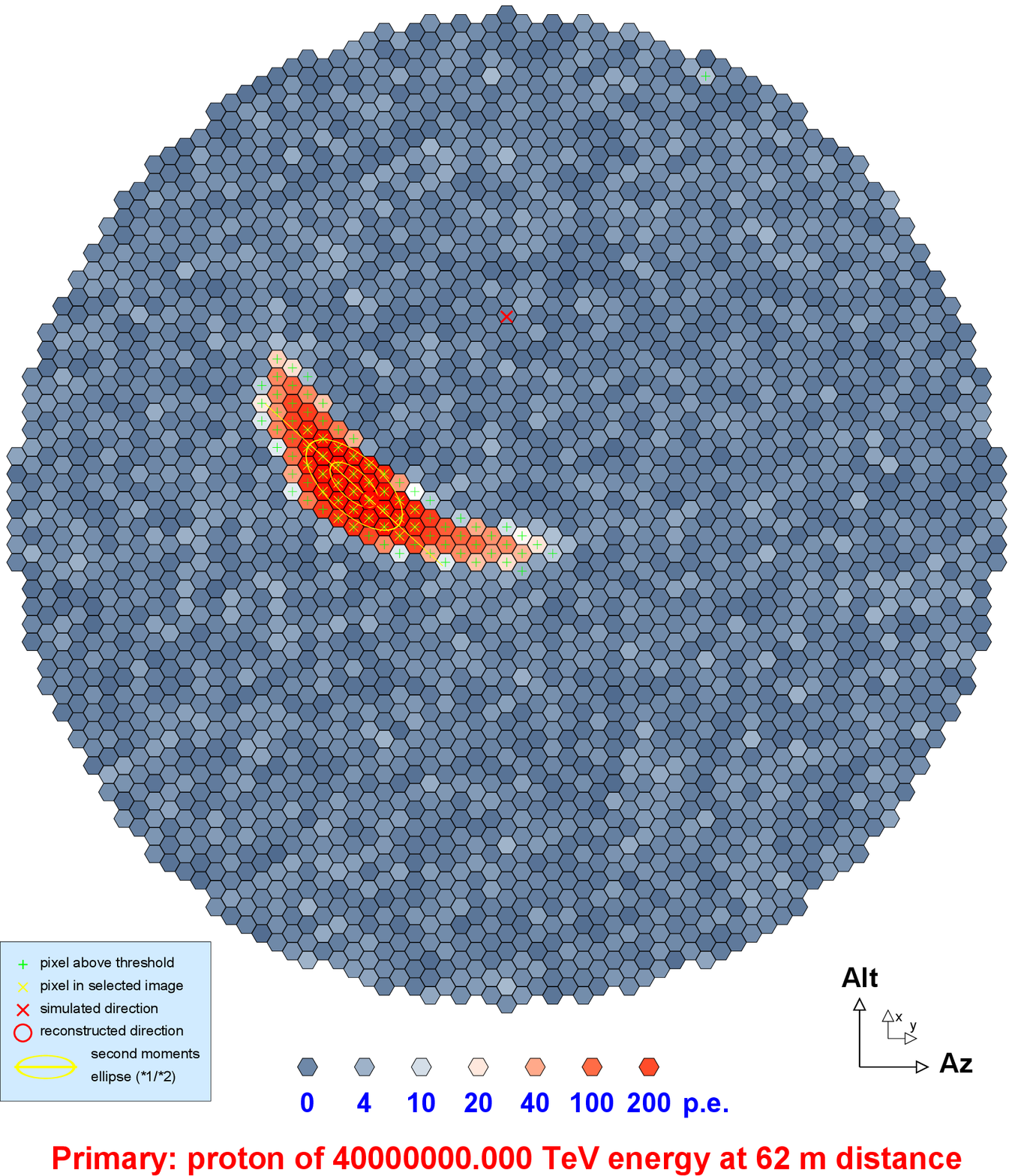}
  \label{fig:sfig5}
\end{subfigure}
\begin{subfigure}{.49\textwidth}
  \centering
  \includegraphics[width=.5\textwidth]{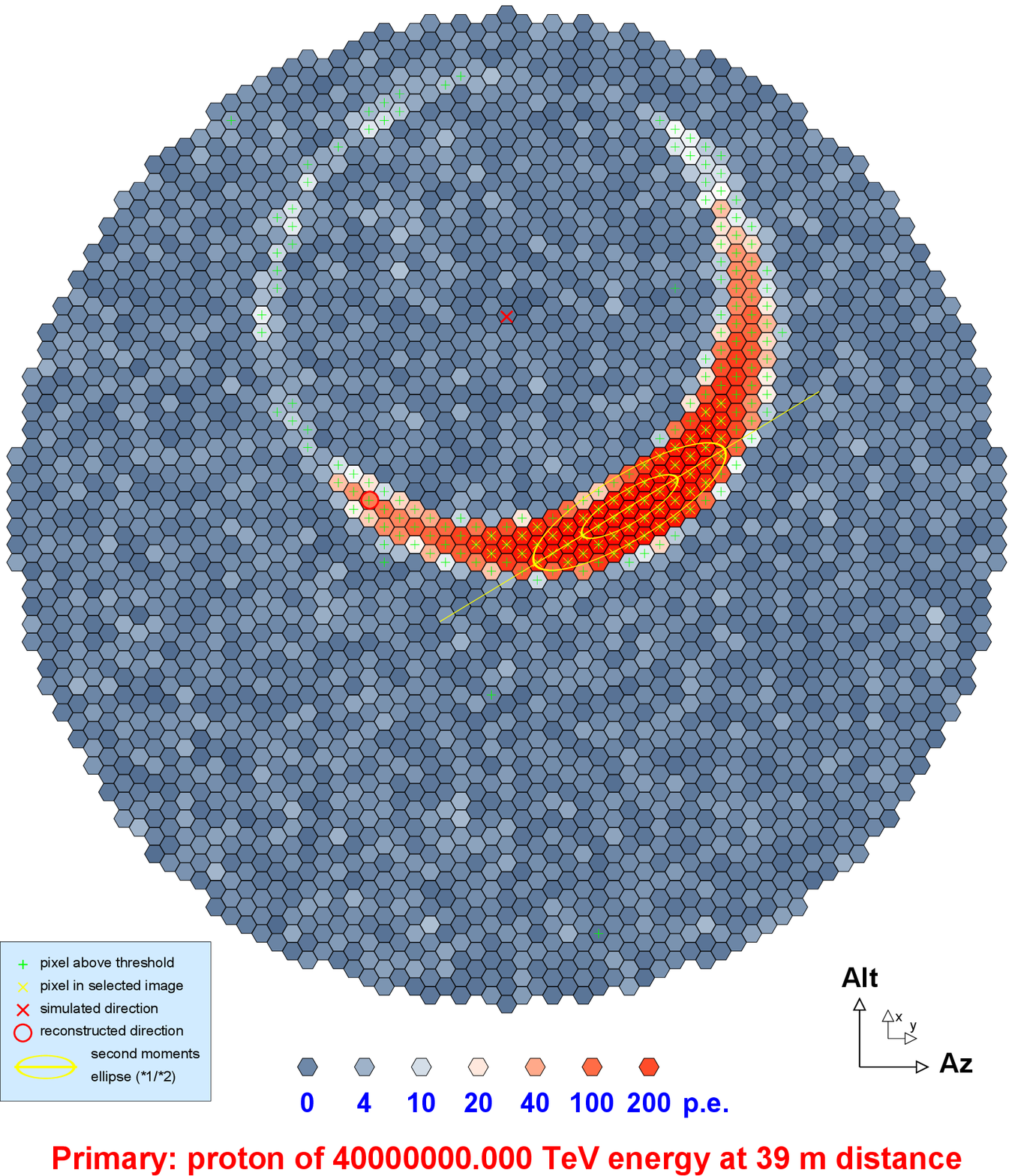}
  \label{fig:sfig6}
\end{subfigure}
\begin{subfigure}{.49\textwidth}
  \centering
  \includegraphics[width=.5\textwidth]{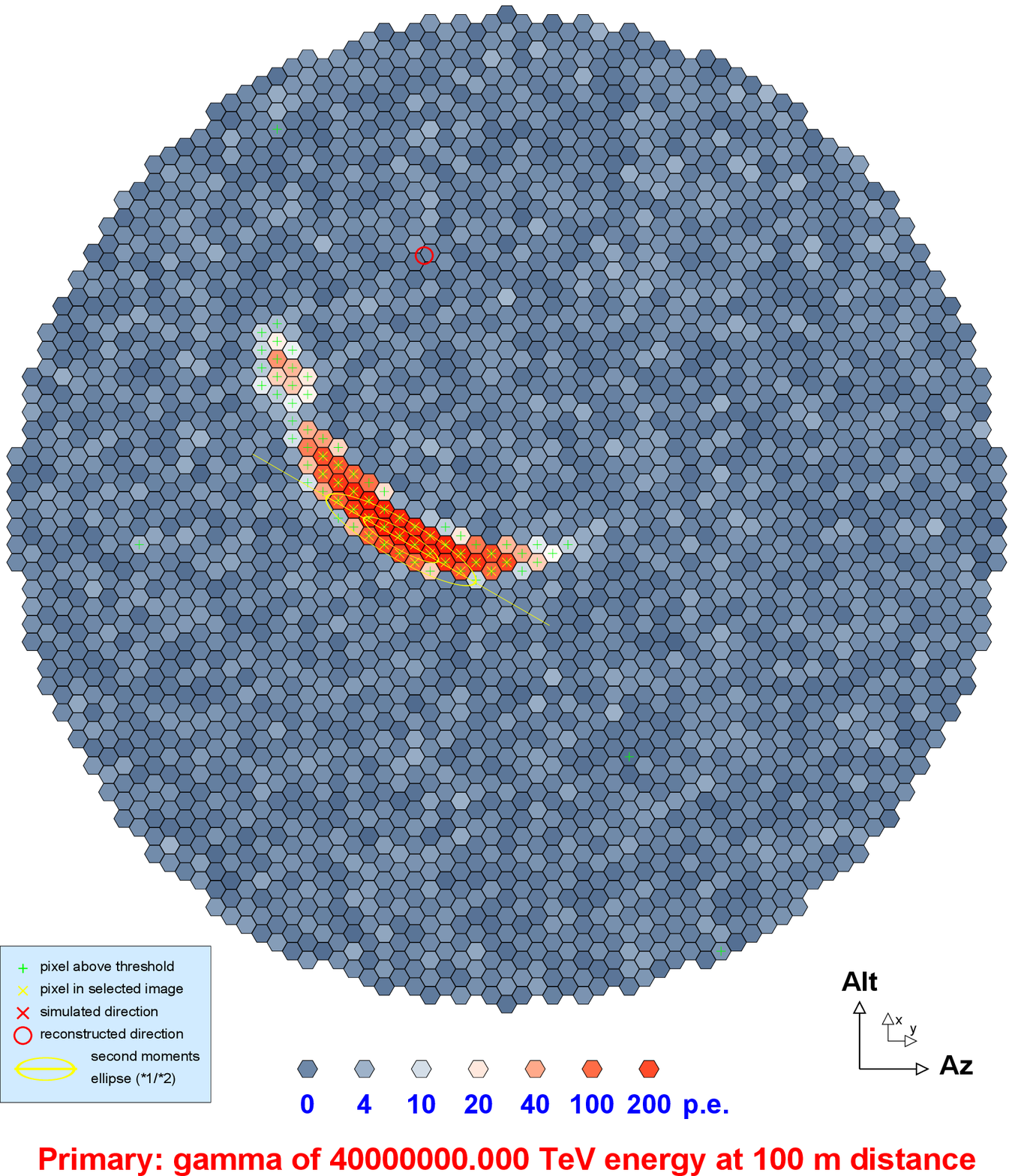}
  \label{fig:sfig7}
\end{subfigure}
\begin{subfigure}{.49\textwidth}
  \centering
  \includegraphics[width=.5\textwidth]{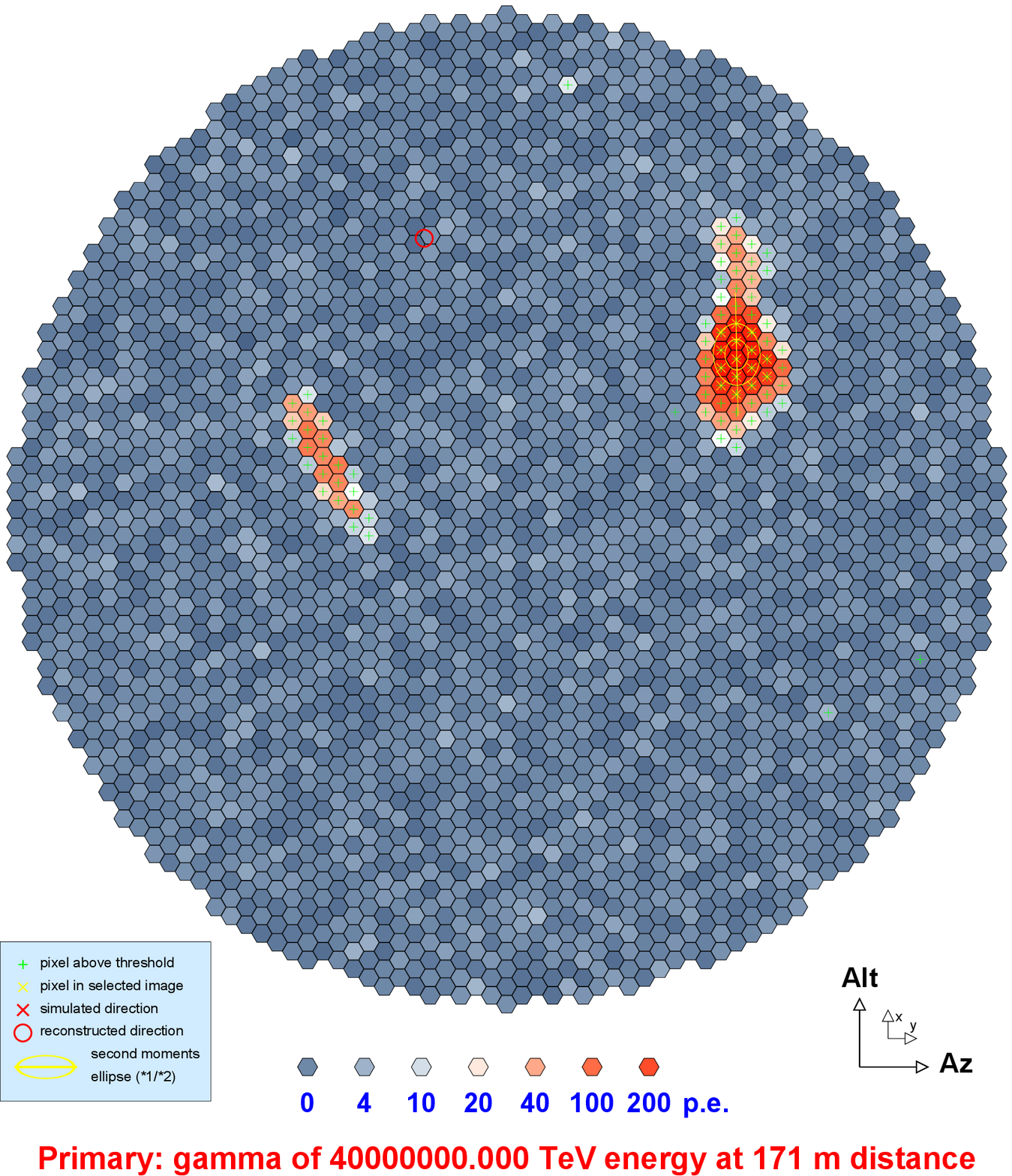}
  \label{fig:sfig8}
\end{subfigure}
\caption{Example of camera images for proton (top) and SPS (bottom) primaries with $E=40$ EeV, $\theta=80^{\circ}$ and $\phi=180^{\circ}$. Sizes in photo-electrons (p.e.) $\rightarrow$ Top-left: 167701 p.e.; Top-right: 313362 p.e.; Bottom-left: 22354 p.e.; Bottom-right: 28491 p.e.}
\label{Figure 4}
\vspace{-0.4cm}
\end{figure}

In case of typical showers with primary energy in GeV-TeV range observed by IACTs at large zenith angles, the Cherenkov light has to undergo a long optical path, due to a thicker layer of the atmosphere. The shower maximum is located far (a few hundred kms) from the observatory and the photon density at the mirrors decreases. This reduces the efficiency compared to lower zenith angles, especially at low energies. Images on the camera will be dimmer and smaller in size. However, at energies of our interest (EeV-ZeV) the so-called gamma-hadron separation is recovered again. Figure 4 shows the images formed on the cameras for proton and SPS primaries. A common feature is the presence of muon rings \cite{vacanti94} characterizing the existence of the muon component shown in Figure 3 for both cases. Because SPS-induced air showers are muon-poor and are initiated higher in the atmosphere, the amount of Cherenkov photons forming these rings will be smaller than in the case of proton primaries and the average number of muon rings for a defined set of telescopes will be lower. However, we also expect a new class of events in the case of SPS such as for example two showers, as shown on the bottom-right camera of Figure 4.

\begin{figure}[t]
\vspace{-0.7cm}
\centering
  \includegraphics[width=.32\textwidth]{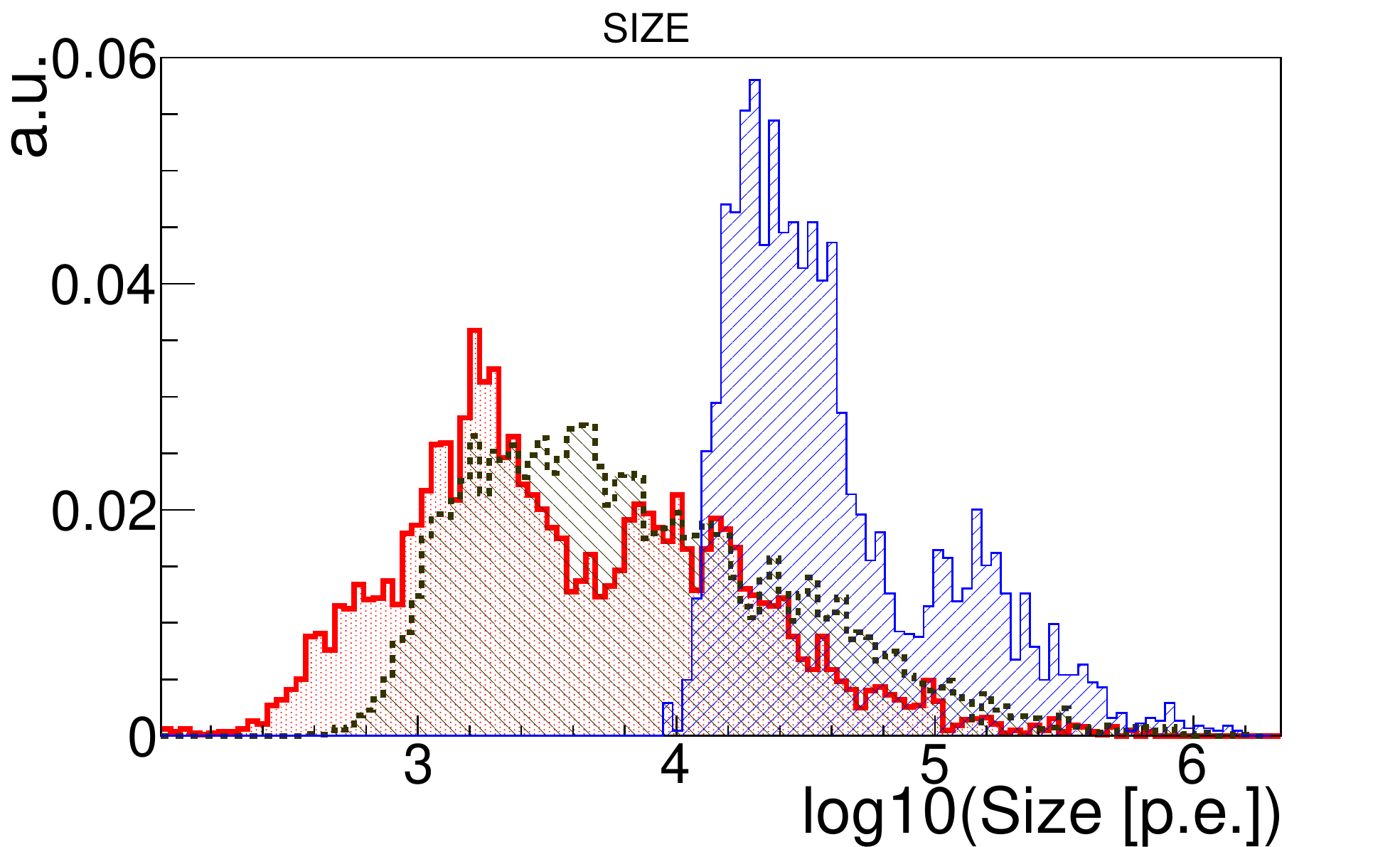}
  \includegraphics[width=.32\textwidth]{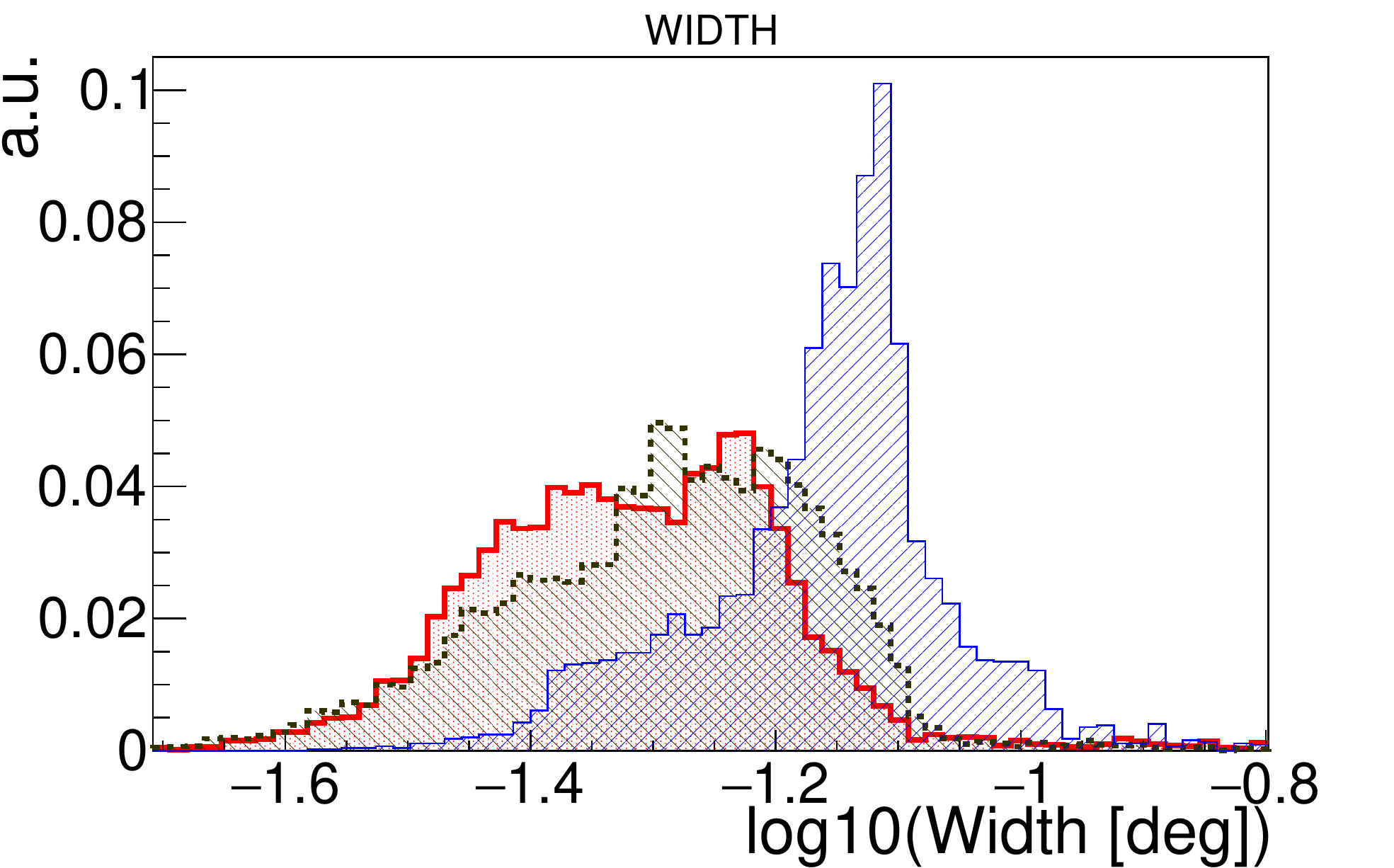}
  \includegraphics[width=.32\textwidth]{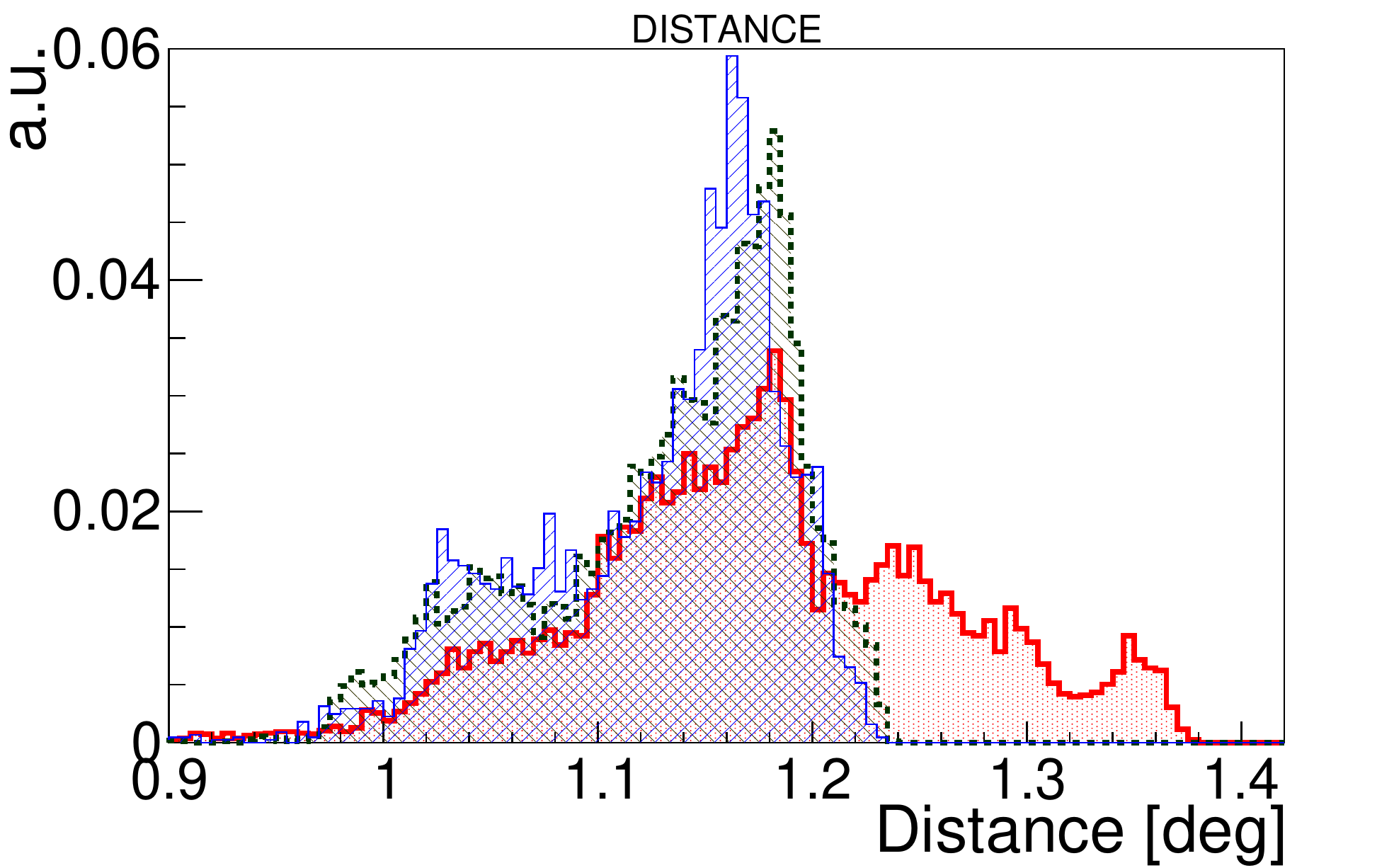}
  
  \medskip
  \includegraphics[width=.32\textwidth]{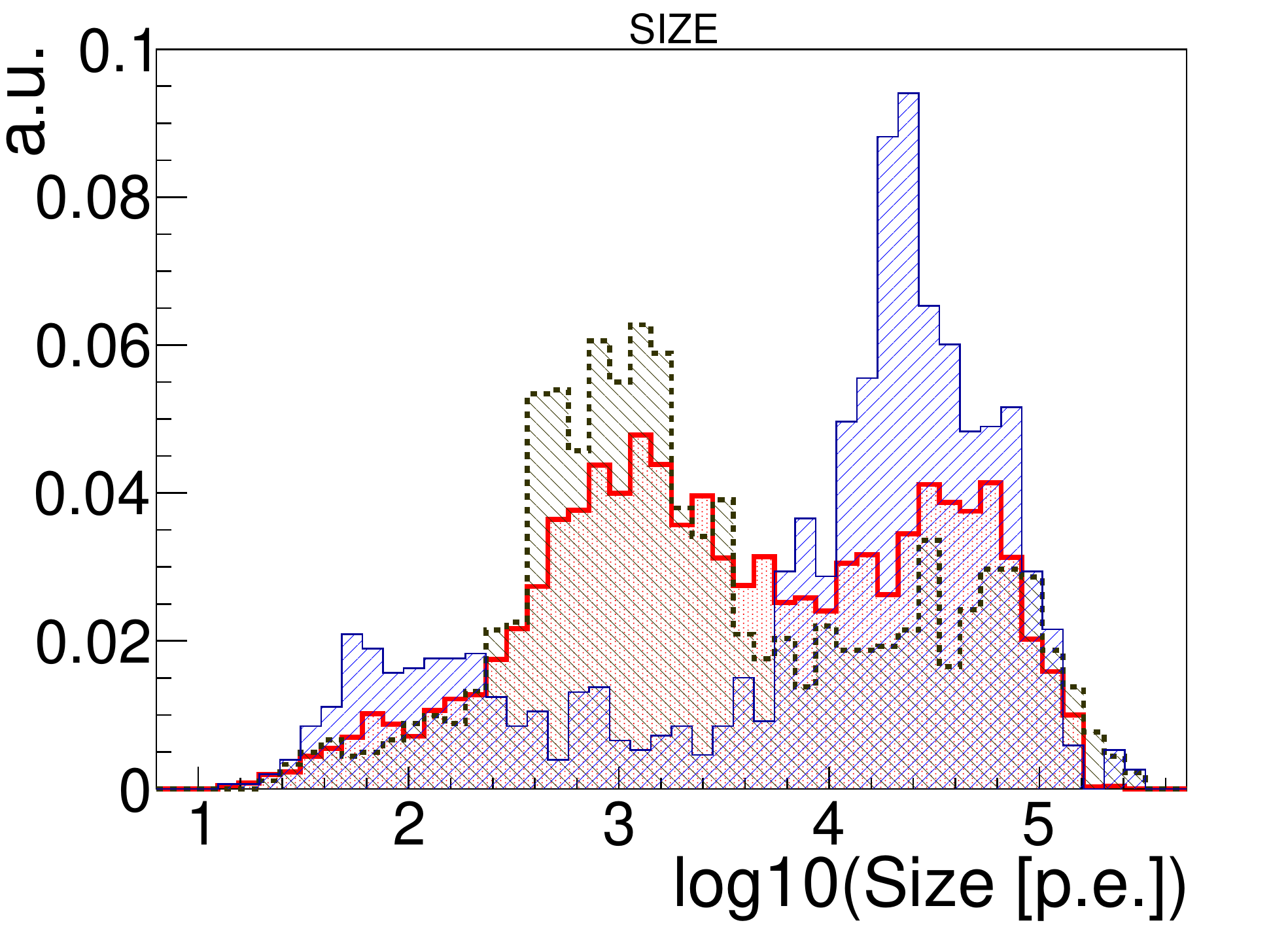}
  \includegraphics[width=.32\textwidth]{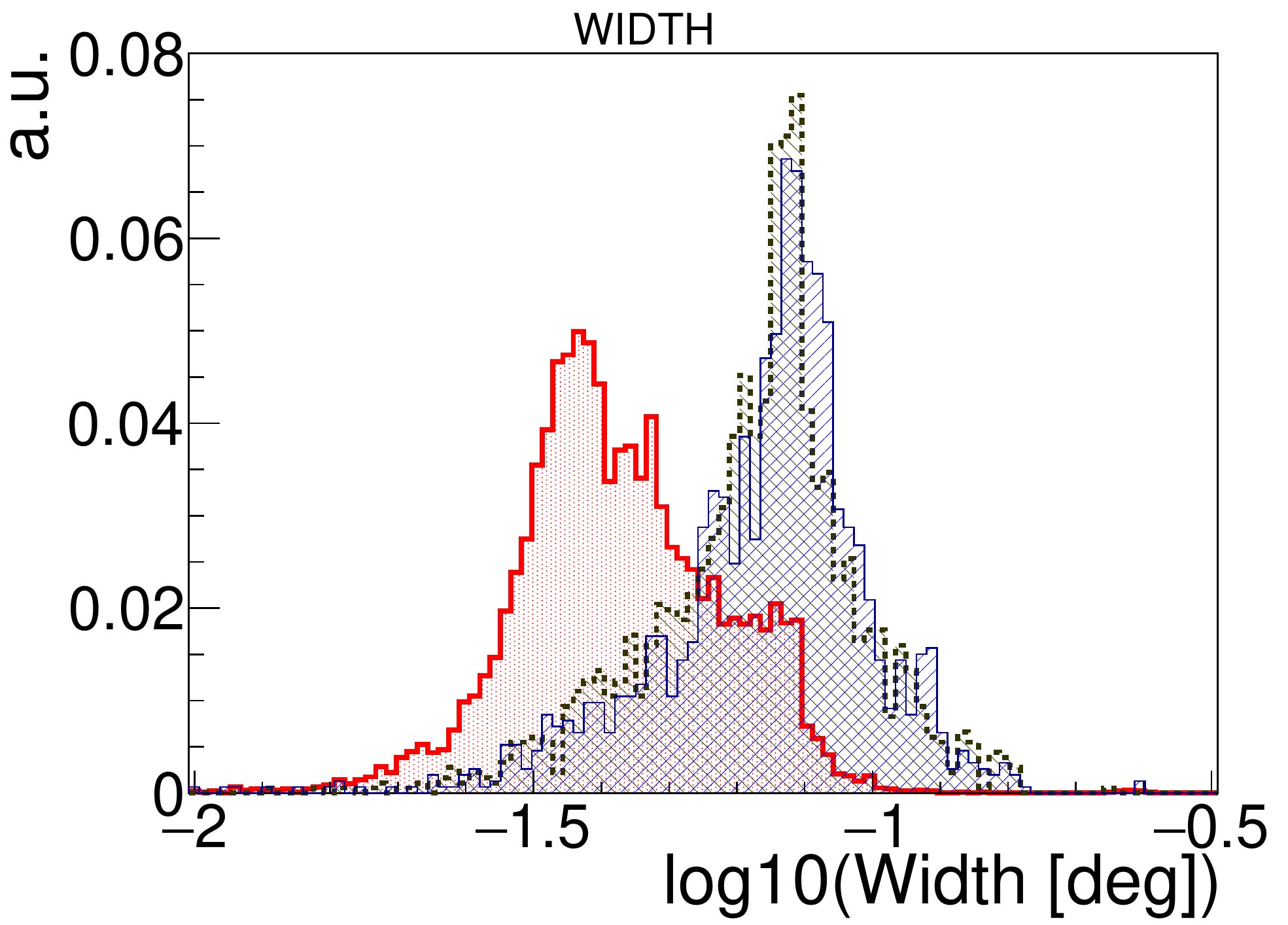}
  \includegraphics[width=.32\textwidth]{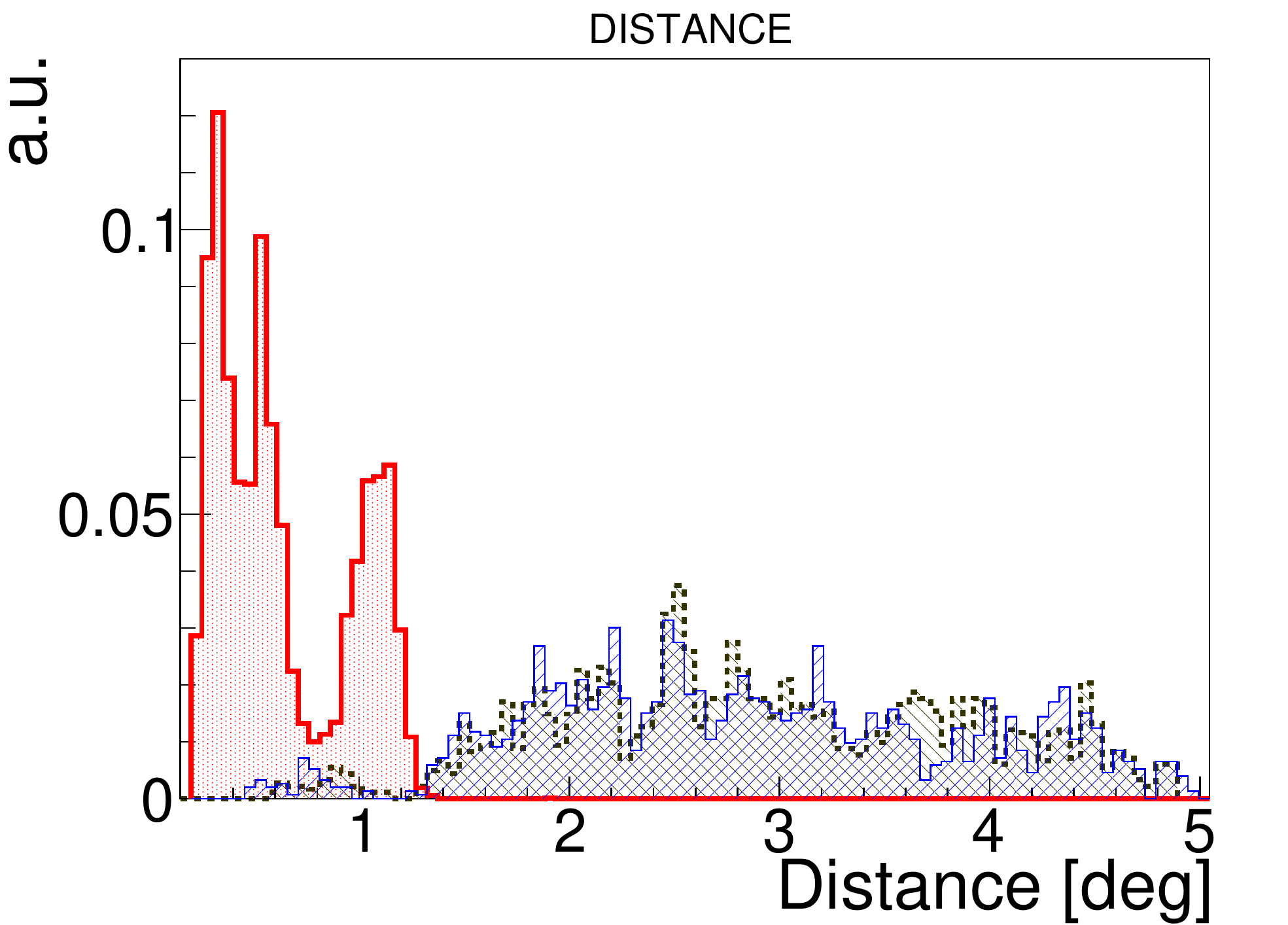}
  
\caption{Normalized distribution of Hillas parameters for proton (thin blue line), photon (dashed brown line) and SPS-induced (thick red line) showers, zenith angle $\theta=80^{\circ}$, azimuth angle $\phi=180^{\circ}$ and $E=40$ EeV. Top panels are for impact parameter $R_{imp}=50$ m and bottom panels for $R_{imp}=1300$ m. \textit{Left panels}: Size. \textit{Center panels}: Width. \textit{Right panels}: Distance.}
\label{Figure 5}
\vspace{-0.5cm}
\end{figure}

Each simulated event recorded and calibrated consists of a number of photoelectrons (p.e.) collected by each pixel in the camera while the trigger gate is opened. The standard trigger configuration requires at least three connected pixels with a signal above the discriminator threshold. However, most of the camera pixels collected light not from the Cherenkov shower but from background. To eliminate the background contribution an image cleaning is performed. The resulting cleaned shower image contains only the pixels considered to have physical information on the shower development. The cleaned camera image is characterized by a set of image parameters introduced by M. Hillas in \cite{hillas85}. These parameters provide a geometrical description of the images of showers and are used to infer the energy of the primary particle, its arrival direction and to distinguish between gamma-ray showers and hadronic showers. It is interesting to study these parameters also in the case of SPS showers. Figure 5 shows three of these parameters for different primaries (SPS, proton and photon) with different impact parameters (50 m and 1300 m). The \textit{Size} parameter measures the total amount of detected light (in p.e.) in all camera pixels, and is correlated with the primary energy of the shower. For both impact parameters, the size-distribution for the proton is shifted towards larger values compared to SPS and photon primary as a consequence of the larger amount of muons capable of emitting Cherenkov photons (Figure 3 - left panel). Images of SPS-induced showers are typically smaller than in the case of proton-induced showers. Therefore, the average value for the \textit{Width} (parameter corresponding to the r.m.s spread of light in direction perpendicular to the image axis) is shifted towards lower values. The angular distance between the center of the shower image and the camera center is called the \textit{Distance} parameter. It is correlated with the angle between the shower and the telescope axis, and for larger zenith angles it decreases due to larger detector-to-shower distance. However, this parameter can also increase when the detector-to-shower distance becomes smaller for fixed zenith angles as one can observe when comparing the top right and bottom right panels of Figure 5 for SPS primary which is considered as a point-like source. However, this behavior is not observed in the case of photon and proton primaries. For any impact distances studied in this work (up to 4 km), the Hillas distributions are different, thus give a possibility to distinguish SPS-induced showers or photons showers from the background of nearly-horizontal hadronic showers.

The expected event rate of preshowers (types A and B) can be estimated from prediction of SHDM model, see Figure 1 - left panel. Assuming integrated gamma-ray flux from SHDM model at level $\phi(E>E_{0})\simeq0.03 \mbox{ } \mathrm{km^{-2}year^{-1}sr^{-1}}$  above $E_{0}=40$ EeV and La Palma array collection area of about $A=\pi R_{imp}^{2}= 50 \mbox{ } \mathrm{km^2}$ at zenith angle $\theta=80^{\circ}$ and $R_{imp}=4$ km, the expected event rate is $N=\phi(E>E_{0})\times A\times \Delta T \times4\pi=0.01$, considering the typical observation time of IACTs of around $\Delta T = 5$ hours and where factor $\mathrm{4\pi}$ is used to convert diffuse flux to point source.

The precise calculation of the sensitivity for SPS is beyond the scope of this paper but will be subject of the future studies. However, we can give a preliminary estimate based on the method shown in \cite{neronov16} (see Eq.7 and Figure 12), i.e. considering 19 telescopes of future La Palma site, and the aperture for SPS of 40 EeV at level $A'=A\times19=950\mbox{ }\mathrm{km^{2}}$ (in extended observation mode), the sensitivity will be approximately $S=1.33\times10^{-9}\mbox{ } \mathrm{GeVcm^{-2}s^{-1}sr^{-1}}$. The calculated sensitivity will be about 20 times lower than what was presented in \cite{neronov16} for $E_{0}=40$ EeV. However, for C and D-type SPS, we can, in principle, put the limit at lower energies, filling the gap between photon limits from KASCADE experiment and gamma limits from the Pierre Auger Observatory.
\vspace{-0.4cm}
\section{Conclusion}
\vspace{-0.3cm}
In this paper, we have shown that next-generation Cherenkov telescopes such as CTA could potentially distinguish SPS (or photon induced showers) from CRs if large zenith angle are analyzed. Taking into account sensitivity estimate, the observation of SPS events seems to be challenging but not impossible to achieve. In addition, a multi-variate analysis based on Hillas parameters and longitudinal profiles could lead to a better discrimination between SPSs and CRs looking at nearly-horizontal directions.
\vspace{-0.4cm}
\acknowledgments
\vspace{-0.2cm}
This research has been supported in part by PLGrid Infrastructure. We warmly thank the staff at ACC Cyfronet AGH-UST, for their always helpful supercomputing support.


\vspace{-0.3cm}
\begin{thebibliography}{99}
\vspace{-0.3cm}
\bibitem{chung99}
D. J. Chung, et al.,
\emph{Phys. Rev.} {\bf D59} (1999), 023501
[{\tt hep-ph/9802238}]

\bibitem{berezinsky97}
V. Berezinsky, et al.,
\emph{Phys. Rev. Lett} {\bf 79} (1997), 4302
[{\tt astro-ph/9708217}]

\bibitem{rubtsov06}
G. I. Rubtsov, et al.,
\emph{Phys. Rev.} {\bf D73} (2006), 063009
[{\tt astro-ph/0601449}]

\bibitem{niechciol17}
M. Niechciol for the Pierre Auger Collab.,
\emph{Proc. of 35th ICRC} (2017) 

\bibitem{homola17}
P. Homola for the CREDO Collab.,
\emph{CERN - Photon 2017 conf.}, Proceedings in prep

\bibitem{mcbreen81}
B. McBreen, C. J. Lambert,
\emph{Phys. Rev.} {\bf D24} (1981), 2536

\bibitem{homola05}
P. Homola et al.,
\emph{Computer Physics Communications} {\bf 184} (2005), 1468
[{\tt astro-ph/0311442}]

\bibitem{cta11}
CTA Consortium,
\emph{Experimental Astronomy} {\bf 32} (2011), 193
[{\tt astro-ph/1008.3703}]

\bibitem{ctasite}
https://www.cta-observatory.org/about/array-locations/la-palma/

\bibitem{neronov16}
A. Neronov et al.,
\emph{Phys. Rev} {\bf D94} (2016) 123018, 
[{\tt astro-ph/1610.01794}]

\bibitem{heck98}
D. Heck, et al.,
\emph{FZKA Report} {\bf 6019} (1998)

\bibitem{ostapchenko06}
S. Ostapchenko,
\emph{Nuc. Phys. B (Proc. Suppl.)} {\bf 151} (2006), 143
[{\tt hep-ph/0412332}]

\bibitem{bass98}
S. A. Bass,
\emph{Prog. in Part. and Nucl. Phys.} {\bf 41} (1998), 255
[{\tt nucl-th/9803035}]

\bibitem{bernlohr08}
K. Bernl\"ohr,
\emph{Astropart. Phys.} {\bf 30} (2008), 149
[{\tt astro-ph/0808.2253}]

\bibitem{cta13}
K. Bernl\"ohr et al. for CTA Consortium,
\emph{Astropart. Phys.} {\bf 43} (2013), 171
[{\tt astro-ph/1210.3503}]

\bibitem{lpm53}
L. D. Landau et al.,
\emph{Dokl. Akad. Nauk SSSR} {\bf 92} (1953), 535 and 735

\bibitem{risse06}
M. Risse et am.,
\emph{Czechoslovak Journal of Physics} {\bf 56} (2006), A327
[{\tt astro-ph/0512434}]

\bibitem{vacanti94}
G. Vacanti et al.,
\emph{Astropart. Phys.} {\bf 2} (1994), 1

\bibitem{hillas85}
A. M. Hillas,
\emph{Proc. of 19th ICRC} {\bf 3} (1985), 445


\end{thebibliography}
\end{document}